%% file: main.tex
\documentclass[conference, 10pt]{IEEEtran}
\IEEEoverridecommandlockouts
\usepackage{cite}
\usepackage{amsmath,amssymb,amsfonts}
\usepackage{graphicx}
\usepackage{textcomp}
\usepackage{xcolor}
\usepackage{pgfplots}
\pgfplotsset{compat=1.18}
\usepackage{subcaption} 
\usepackage{stfloats}
\usepackage{placeins}
\usepackage{url}

\def\BibTeX{{\rm B\kern-.05em{\sc i\kern-.025em b}\kern-.08em
    T\kern-.1667em\lower.7ex\hbox{E}\kern-.125emX}}
\input{macros}

\IEEEoverridecommandlockouts\IEEEpubid{\makebox[\columnwidth]{979-8-3195-0662-7/26/\$31.00 $\copyright$2026 IEEE \hfill}\hspace{\columnsep}\makebox[\columnwidth]{ }}
\begin{document}

\title{SRAN: Scaling NDN Routing via Map-and-Attach}
\author{
\IEEEauthorblockN{
Tianyuan Yu,
Sirapop Theeranantachai,
and Lixia Zhang
}
\IEEEauthorblockA{
Department of Computer Science\\
University of California, Los Angeles\\
Los Angeles, CA, USA\\
\{tianyuan, stheera, lixia\}@cs.ucla.edu
}
}
\maketitle

\begin{abstract}
Network routing scalability becomes difficult when forwarding state scales with an external identifier space rather than network topology. Named Data Networking (NDN) faces this challenge acutely because routing directly on application name prefixes ties forwarding state to an unbounded namespace. 
This paper presents SRAN, an intra-domain NDN routing architecture that applies the routing-scalability principle underlying Map-and-Encap through Map-and-Attach: application prefixes are mapped to egress routers, and the resulting mapping information is attached to the original Interest. 
This enables network routing and forwarding to operate on topological identifiers while keeping NDN Interests intact and preserving native NDN communication semantics, including Interest/Data exchange, in-network caching, and data-centric security. SRAN further uses the same prefix-to-egress mapping to realize Bit Index Explicit Replication (BIER) for scalable NDN Interest multicast. 
SRAN leverages existing NDN mechanisms to securely maintain prefix-to-router mappings among user-facing routers without introducing new protocols.
Evaluation on Rocketfuel topologies shows that routers' forwarding state scales with network topology rather than application-prefix count, while prefix updates are disseminated in real time with low communication overhead.

\end{abstract}


\input{intro}
\input{back}
\input{design}
\input{eval}
\input{related}
\input{discuss}
\input{conclude}

\section*{Acknowledgment}
We thank our shepherd and the anonymous reviewers for their constructive feedback, which significantly improved this paper. 
We also thank Lan Wang, Beichuan Zhang, Junxiao Shi, and Davide Pesavento for insightful discussions and suggestions that helped us refine the ideas presented in this work.



\bibliographystyle{IEEEtran}
\bibliography{refs} 

\end{document}

%% file: macros.tex
\usepackage{ifthen}
\usepackage[T1]{fontenc}

\usepackage{booktabs}
\usepackage{multirow}
\usepackage{soul}
\usepackage{amsmath,amssymb}
\usepackage[inline]{enumitem}
\usepackage{algorithm}
\usepackage{algpseudocode}

\providecommand{\REQUIRE}{\Require}
\providecommand{\ENSURE}{\Ensure}
\providecommand{\STATE}{\State}
\providecommand{\RETURN}{\Return}

\providecommand{\IF}{\If}
\providecommand{\ELSE}{\Else}
\providecommand{\ENDIF}{\EndIf}
\providecommand{\ELSIF}{\ElsIf}

\usepackage{algpseudocode}
\usepackage{etoolbox, xstring}
\usepackage{xspace}
\DeclareListParser{\doslashlist}{/}
\newcounter{ndnNameComponentCounter}%
\newcommand{\name}[1]{{%
		\setcounter{ndnNameComponentCounter}{0}%
		\renewcommand{\do}[1]{{%
				\ifnumgreater{\value{ndnNameComponentCounter}}{0}{\allowbreak/}{}%
				\ifnumodd{\value{ndnNameComponentCounter}}{}{}%
				\detokenize{##1}}%
			\stepcounter{ndnNameComponentCounter}}%
		``{\fontfamily{cmtt}\small\selectfont\IfBeginWith{#1}{/}{/}{}\doslashlist{#1}}''%
}}

\newboolean{showcomments}
\setboolean{showcomments}{true}
\ifthenelse{\boolean{showcomments}}
{ \newcommand{\mynote}[3]{
    \protect\fbox{\bfseries\sffamily\scriptsize#1}
    {\small$\blacktriangleright$\textsf{\emph{\color{#3}{#2}}}$\blacktriangleleft$}}}
{ \newcommand{\mynote}[3]{}}

\usepackage{xcolor}

\newcommand{\ie}{\textit{i.e.,}\@\xspace}

\definecolor{verylightgray}{gray}{0.8}

\usepackage[normalem]{ulem}

\definecolor{revisionblue}{RGB}{0,82,155}
\DeclareRobustCommand{\rev}[1]{{\color{revisionblue}\sffamily #1}}

%% file: intro.tex
\section{Introduction}
\label{sec:intro}

Named Data Networking (NDN)~\cite{zhang2014named} lets producers announce data name prefixes to the routing system, and
consumer applications issue Interest packets for named, secured data. Routers forward those Interests toward producers, data delivery benefits from in-network caching, and the network can quickly adapt to failures and support load balancing through its stateful forwarding plane.
Because data producers announce their name prefixes to the routing system, NDN faces a more severe routing scalability challenge than IP: its FIB can grow with an unbounded application namespace and experience prefix churn driven by application dynamics.
Moreover, NDN supports multiparty applications via its transport protocol Sync, which relies on network multicast forwarding to deliver Sync Interests to participants in the same application~\cite{zhu2013chronosync,zhang2016psync,moll2020svs,moll2022sok}.
Unfortunately, traditional network multicast solutions cause routers' forwarding state to grow with the number of multicast groups.

The two problems share the same root cause: network routing and forwarding state is driven by logical identifiers---application name prefixes for unicast and multicast groups for multicast---whose number is determined by applications rather than network topology.
For example, a Sync-based collaborative application~\cite{moll2021brokerless,yu2024ndnworkspace} registers unicast prefixes for each user and group multicast prefixes.
Routing each application prefix and multicast group directly causes routing and forwarding state to grow with application dynamics outside network operators' control.

Our recent study of four decades of IP routing scalability efforts identifies a recurring architectural principle underlying Map-and-Encap solutions: map external identifiers to their topological attachment points, thereby allowing routing and forwarding to operate on topology-derived identifiers~\cite{yu2026mapencapbier}.
Conventional Map-and-Encap realizes this principle by encapsulating the original packet with an outer header identifying the mapped topological destination. The same study also observes that BIER applies this mapping principle to multicast by mapping a multicast group's destinations to the topology-bound egress routers through which they are reachable, and encoding those router identifiers in a BitString.

However, applying this mapping principle to NDN presents a unique challenge.
Conventional encapsulation would turn an Interest into an opaque payload addressed to a topological endpoint, preventing intermediate NDN routers from aggregating Interests with the same application name in the Pending Interest Table (PIT) and from using those names to look up the in-network cache.

This paper introduces \textbf{S}calable \textbf{R}outing \textbf{A}rchitecture for \textbf{N}DN \textbf{(SRAN)}.
SRAN supports scalable unicast and multicast routing and forwarding while preserving the native data-centric communication model by realizing the same topology-mapping principle through Map-and-Attach rather than encapsulation. It maps the Interest name to an egress router but carries the mapping result in NDNLP~\cite{ndnlpv2} alongside the intact Interest.
Specifically, we make three contributions:

\begin{itemize}
\item We design SRAN, an intra-domain scalable routing architecture for NDN that realizes the topology-mapping principle through Map-and-Attach: SRAN maps application prefixes to egress routers and carries the mapping information in NDNLP~\cite{ndnlpv2}, leaving NDN Interests intact for native data-centric forwarding.

\item We build a distributed prefix-to-router mapping system using existing NDN mechanisms, including dataset synchronization and data-centric security, without introducing new protocols.

\item Experiments with our prototype show that SRAN keeps routers' forwarding state bounded by network topology rather than application-prefix scale, while achieving real-time prefix-to-router mapping updates with low dissemination overhead.
\end{itemize} 

The remainder of this paper is organized as follows.
Section~\ref{sec:back} provides background information on NDN and BIER.
Section~\ref{sec:design} presents the SRAN design, including Map-and-Attach forwarding, multicast support, prefix-to-router mapping maintenance, and routing security.
Section~\ref{sec:eval} describes our prototype implementation and evaluation results.
Section~\ref{sec:related} summarizes the related work and Section~\ref{sec:discuss} clarifies the architectural contribution and discusses the applicability of the same routing-scalability principle to inter-domain routing.
We conclude the paper in Section~\ref{sec:conclude}.

%% file: back.tex
\section{Background}
\label{sec:back}

\subsection{Named Data Networking}
\label{sec:back:ndn}
Named Data Networking (NDN)~\cite{zhang2014named,2018ndn-intro} shifts network communication from host-to-host delivery to data retrieval, where applications fetch named and secured Data.

\noindent\textbf{Named Secured Data:}
NDN applications assign each data unit a semantically meaningful name.
These names serve as the key for application logic, transport protocols, and network-layer routing.
Every Data packet is cryptographically signed by its producer, binding the name to the content and enabling data-centric security based on semantic naming~\cite{yu2015schematizing,yu2023new,zhang2018nac}.

\begin{figure}[t]
    \centering
    \includegraphics[width=0.45\textwidth,
    trim={0 0 9cm 0},
    clip]{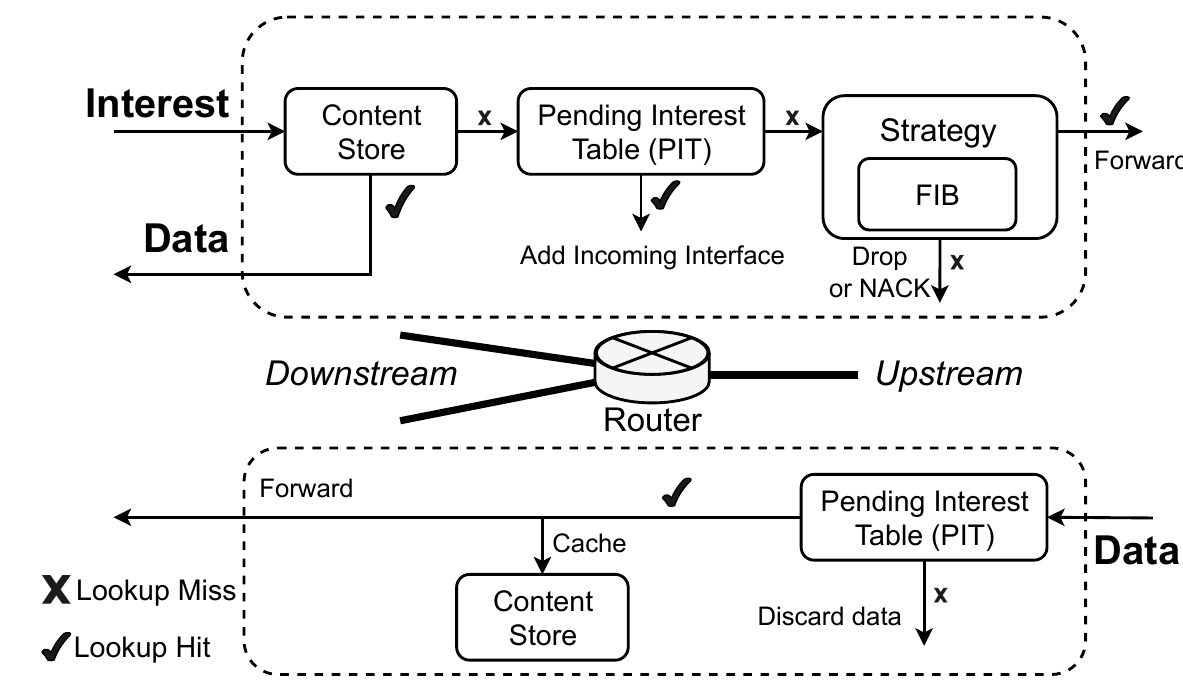}
    \caption{NDN's data-centric forwarding pipeline. An Interest may be satisfied by cached Data, aggregated in the PIT, or forwarded by name-based FIB lookup.}
    \label{fig:ndn101}
\end{figure}

\noindent\textbf{Data-Centric Forwarding:} NDN provides data-centric delivery with two packet types: \textit{Interests} and \textit{Data}.
As illustrated in Fig.~\ref{fig:ndn101}, a router processes an incoming Interest through a sequential lookup: it is first satisfied by the \textit{Content Store} (CS) if cached, aggregated in the \textit{Pending Interest Table} (PIT) if a matching data request is already outstanding, or otherwise forwarded via the \textit{Forwarding Information Base} (FIB) using a name-based strategy.
This stateful pipeline ensures that returning Data follows the reverse path of the Interest, enabling in-network data request aggregation.

\noindent\textbf{Routing Scalability:}
The route computation algorithm itself is independent of the application namespace since it computes paths through the network topology.
However, the routing system needs to associate each reachable application prefix with these routes and install the corresponding FIB entries.
Because application prefixes are created independently and their number grows with the number of producers, routing and forwarding state scales with the potentially unbounded application namespace.

\noindent\textbf{Link Adaptation Layer:}
The NDNLP protocol~\cite{ndnlpv2} provides a link adaptation layer between the network and link layers, allowing additional information needed by forwarders to be \emph{attached} to Interest or Data packets as header fields without modifying the network-layer packets~\cite{schneider2016pcon,song2022effective}.

\noindent\textbf{Dataset Synchronization:}
Multiparty NDN applications~\cite{hoque2013nlsr,gusev2015ndnrtc,yu2024ndnworkspace} typically use Sync protocols~\cite{zhu2013chronosync, zhang2016psync, moll2020svs, moll2022sok} to maintain a shared state among group entities via a cumulative state vector.
When a producer generates data, it \emph{multicasts} a Sync Interest carrying the updated state vector to the group, allowing all participants to discover new data names promptly.

\subsection{Bit Index Explicit Replication (BIER)}
\label{sec:back:bier}


BIER~\cite{rfc8279} uses mapping information to convert multicast destinations into a bitstring.
For each multicast flow, a multicast flow overlay provides the Bit-Forwarding Ingress Router (BFIR) with the set of egress routers (BFERs) that must receive the packet.
BIER itself does not prescribe how this mapping is learned~\cite{rfc8279}, and one option is using iBGP in MVPN deployments~\cite{rfc8556}. 
Each BFER is assigned a BFR-ID, while the routing underlay distributes BFR-ID reachability and constructs a Bit Index Forwarding Table (BIFT). 
The BFIR maps the destination BFERs to their BFR-IDs, encodes them as a BitString, and encapsulates the original packet with a BIER header. 
Transit routers then replicate and forward the packet using only the BitString and BIFT, without maintaining per-flow multicast state or inspecting the encapsulated packet.


%% file: design.tex
\section{SRAN Design}
\label{sec:design}
This section presents the SRAN design in an NDN network.
We first present a system overview, followed by the forwarding details of Map-and-Attach together with the BIER-based stateless multicast support.
Afterwards, we explain the synchronization of the mapping table and its security support.

\subsection{System Overview}
\label{sec:design:design}
SRAN's design goal is to map the application namespace onto the network topology for scalable routing and forwarding without changing NDN's data-centric communication model.
SRAN meets this goal by keeping NDN Interest packets intact, enabling the network to perform normal NDN Interest/Data exchange, PIT processing, and in-network caching. 
SRAN therefore does not encapsulate an Interest in another network-layer packet.
Instead, it maps the Interest name to a router and \emph{attaches} the resulting mapping information alongside the original Interest as forwarding metadata in NDNLP.

SRAN assumes that all routers run an NDN routing protocol, such as NLSR~\cite{afanasyev2015nlsr} or NDN Distance Vector~\cite{patil2024poster}, to learn router-level reachability.  
This router-level network routing is independent of the application namespace and of the mechanism SRAN uses to obtain and maintain prefix-to-router mappings.

When an Interest first enters the network, the user-facing ingress router maps the Interest name to one or more egress routers $R_e$, \ie routers through which producers of the requested data can be reached.
SRAN maintains this mapping in the Prefix Egress Table (PET) (Section~\ref{sec:design:pet}).
For unicast delivery, the router selects an egress router $R_e$ from the corresponding PET entry and places its identifier in the NDNLP header associated with the Interest.
The original Interest and its $R_e$ then travel together through the network.
Each router along the path looks up the $R_e$ in its FIB and forwards the Interest toward the resulting next hop.
Once the Interest reaches the $R_e$, it forwards the Interest to the locally attached producer based on the Interest's data name.  

The same prefix-to-router mapping also supports Interest multicast.
Each multicast prefix announcement carries a special flag to distinguish itself from unicast prefixes. 
Upon receiving a multicast Interest of name $N$, the ingress router finds all egress routers attached to name $N$ from PET, and encodes the destination router set as a BIER BitString to be carried in NDNLP (Section~\ref{sec:design:multicast}).  
Routers then use topology-derived BIFT state to replicate and forward the Interest toward the corresponding egress routers, without knowing any information about multicast groups.

SRAN maintains a prefix-to-router mapping table on all user-facing routers.
Applications securely announce their prefixes to the router to which they are attached (Section~\ref{sec:design:sec}).
Each such router records its local prefix reachability in the Prefix State Database (PSD), and the routers synchronize their PSD state (Section~\ref{sec:design:psd}).


\subsection{Forwarding Plane: Prefix Egress Table}
\label{sec:design:pet}

\begin{figure*}[t]
    \centering
    \begin{subfigure}[t]{0.49\textwidth}
        \centering
        \includegraphics[
            trim={0.5cm 0.5cm 0.5cm 0.3cm},
            clip,
            width=\textwidth
        ]{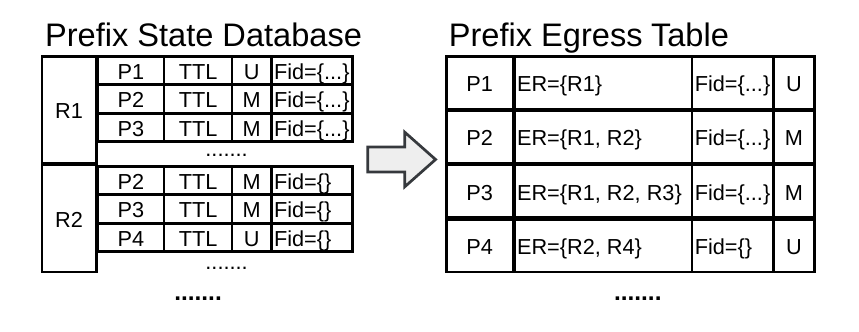}
        \caption{The mapping tables in router R1.}
        \label{fig:psd}
    \end{subfigure}
    \hfill
    \begin{subfigure}[t]{0.49\textwidth}
        \centering
        \includegraphics[
            trim={0.5cm 0.3cm 0.5cm 0.5cm},
            clip,
            width=\textwidth
        ]{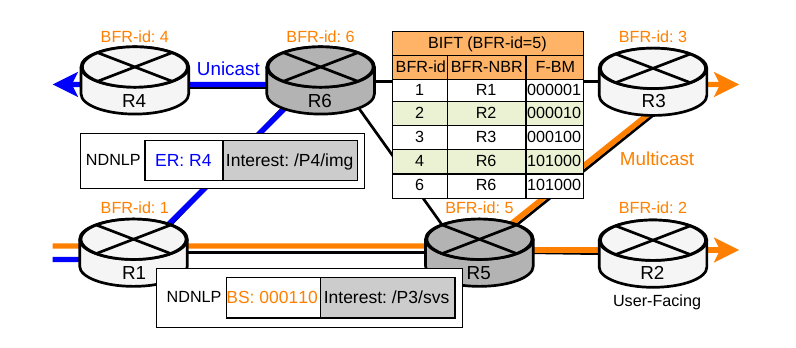}
        \caption{A network forwards unicast and multicast Interests.}
        \label{fig:forward}
    \end{subfigure}
    \caption{Map-and-Attach in SRAN. (a) The Prefix State Database (PSD) aggregates prefix reachability updates from routers to maintain the router-to-prefix mappings, which populate the Prefix Egress Table (PET) that maintains the prefix-to-router mappings. Fids are router-local face identifiers and are not used by remote routers for forwarding. TTL provides soft-state cleanup when an application or user-facing router disappears without an explicit withdrawal, and a live application refreshes its announcement before expiration. (b) Ingress routers consult the PET to map an application prefix to topological locators: unicast Interest packets are attached with a specific Egress Router (ER) in the associated NDNLP header, while multicast Interest packets are attached with a BitString (BS) and forwarded using BIFT entries that store per-face Forwarding Bit Masks (F-BMs).}
    \label{fig:psd-forward}
\end{figure*}

The PET is SRAN’s prefix-to-router mapping table, organized for longest-prefix matching.
Each PET entry maps an application name prefix to a set of egress routers $\mathcal{E}$, a set of local faces $\mathcal{F}$ that are meaningful only at the user-facing router where the application is attached, and a type of delivery (unicast or multicast).
Populated from the PSD, the PET is used directly on the forwarding path.
Fig.~\ref{fig:psd} shows this relationship on router R1, where PSD state is materialized into PET entries.
Alg.~\ref{alg:forwarding} summarizes the router forwarding process.
We walk through the forwarding lifecycle of an Interest using the unicast scenario in Fig.~\ref{fig:forward} as a running example.

\noindent\textbf{Map-and-Attach:}
To bridge the application namespace and physical topology, the ingress router first queries the PET to map the Interest name to $\mathcal{E}$;
here we use \emph{ingress router} to denote the user-facing router at which an Interest enters the SRAN domain; it is a per-Interest role, not a distinct router type.
For example, when ingress router R1 receives an Interest for \name{/P4/img} from an application, it maps $P4$ to R2 and R4 as candidate egress routers.
Assume that R4 has a lower routing cost; R1 therefore chooses R4 as the $R_e$.
In Alg.~\ref{alg:forwarding}, \textsc{BestRoute}$(\mathcal{E})$ denotes this selection.
The ingress router R1 then attaches the Interest from the application within an NDNLP header containing the target $R_e$.
In our running example, R1 adds an NDNLP header with R4 as the Interest $R_e$.

\begin{algorithm}[!t]
\caption{Unicast Interest Forwarding (PIT and CS lookups are omitted)}
\label{alg:forwarding}
\begin{algorithmic}[1]
\REQUIRE Incoming Packet $P$ (incoming Interest $I_o$, or Interest with attached egress-router $I_a$)
\ENSURE $P$ forwarded, locally delivered, or dropped

\IF{$P$ is $I$}
    \STATE $(\mathcal{E}, \mathcal{F}) \gets \text{PET.Lookup}(I_o.\text{name})$
    \IF{$\mathcal{E} \cup \mathcal{F} = \emptyset$}
        \STATE \textbf{drop} $P$
        \RETURN
    \ENDIF
    \IF{$\mathcal{E} = \emptyset$}
        \STATE \textbf{deliver} $I_o$ to $\mathcal{F}$
        \RETURN
    \ENDIF

    \STATE $R_e \gets \text{BestRoute}(\mathcal{E})$
    \STATE $I_a \gets \text{Attach}(I_o, R_e)$
    \STATE $f \gets \text{FIB.Lookup}(R_e)$
    \STATE \textbf{forward} $I_a$ to $f$

\ELSIF{$P$ is $I_a$}
    \STATE $(I_o, R_e) \gets \text{Parse}(P)$
    \IF{$R_e = R_{me}$}
        \STATE $(\_, \mathcal{F}) \gets \text{PET.Lookup}(I_o.\text{name})$
        \STATE \textbf{deliver} $I_o$ to $\mathcal{F}$
    \ELSE
        \STATE $f \gets \text{FIB.Lookup}(R_e)$
        \STATE \textbf{forward} $P$ to $f$
    \ENDIF
\ENDIF
\end{algorithmic}
\end{algorithm}

\noindent\textbf{Forwarding towards the Egress Router:}
After attaching $R_e$ in the NDNLP header, the ingress router then queries the chosen $R_e$ name against its FIB, which maintains only the router reachability state, to find the outgoing face.
In subsequent hops, after the CS lookups and PIT aggregation by Interest names, the routers forward Interests only by their associated NDNLP headers.
Because the FIBs contain no application prefixes and store only router reachability state, these routers perform exact-match lookups on the $R_e$ name.
Consequently, the FIB at every router scales with the physical topology rather than the application namespace.
When the Interest reaches router R6, it matches the R4 name in the NDNLP header against its local FIB and forwards the Interest to R4.
Because PIT aggregation and CS lookups are still based on the original Interest names, the data-centric communication model is preserved.

\noindent\textbf{Egress Delivery:}
The egress router detects that an Interest has reached its destination when the attached $R_e$ matches its own name.
It performs a final local PET lookup to map the Interest name to a specific local face, delivering the Interest packet to the producer application.
Upon receiving the Interest, R4 determines that the $R_e$ in the NDNLP header matches its own name, and directly delivers \name{/P4/img} to the attached producer.
Because Interest forwarding opens PIT entries along the path, the returning Data can be forwarded in the reverse direction without additional support.

\subsection{Supporting Stateless Interest Multicast}
\label{sec:design:multicast}
SRAN supports stateless Interest multicast with BIER forwarding\footnote{SRAN does not remove NDN's normal per-Interest PIT state, which is essential for Data forwarding.}.
BIER requires the ingress router to know which BFR-IDs should receive a packet; in SRAN, the PET natively provides this mapping for name prefixes.
Because a PET entry maps a multicast prefix to its participating egress routers, an ingress router can map an Interest name directly into the egress router set.

In SRAN, each router is configured with a unique BFR-ID, representing a specific bit position in a domain-wide BitString (BS).
Multicast prefixes are explicitly labeled in the PET, and an ingress router identifies multicast Interests and retrieves the destination egress routers directly through a PET query.
The ingress router then encodes those egress routers into a BS and replicates the packet to the appropriate outgoing faces learned from the BIFT lookup.
Each BIFT entry stores a Forwarding Bit Mask (F-BM), the set of BFR-IDs reachable through a given outgoing face.
As illustrated by the multicast scenario in Fig.~\ref{fig:forward}, when an application attached to R1 expresses a multicast Interest \name{/P3/svs}, through PET lookup, R1 identifies \name{/P3} as a multicast prefix and R2 and R3 as destination routers.
It encodes their BFR-IDs in a single BS $000110$ within the NDNLP header and forwards the packet based on the local BIFT.

Routers also process these Interests based on the topology-bound BIFT.
Upon receiving an Interest, a router first checks whether its own BFR-ID is set in the incoming BS, indicating that the Interest has reached one of its destinations.
If the bit is set, the router performs a PET lookup to deliver the Interest to local faces, then clears its own bit from the BS to prevent local loopback.
Afterwards, the router maps the remaining bits in BS to outgoing interfaces using BIFT and replicates the packet once per interface carrying at least one destination.
This approach ensures that BIER forwarding state remains invariant to the number of active multicast groups because BIFT state is tied to router reachability rather than application names.
In Fig.~\ref{fig:forward}, the intermediate router R5 replicates and forwards the Interest to both R2 and R3 based entirely on the $000110$ BitString.

\begin{figure*}[t]
    \centering
    \includegraphics[width=\linewidth]{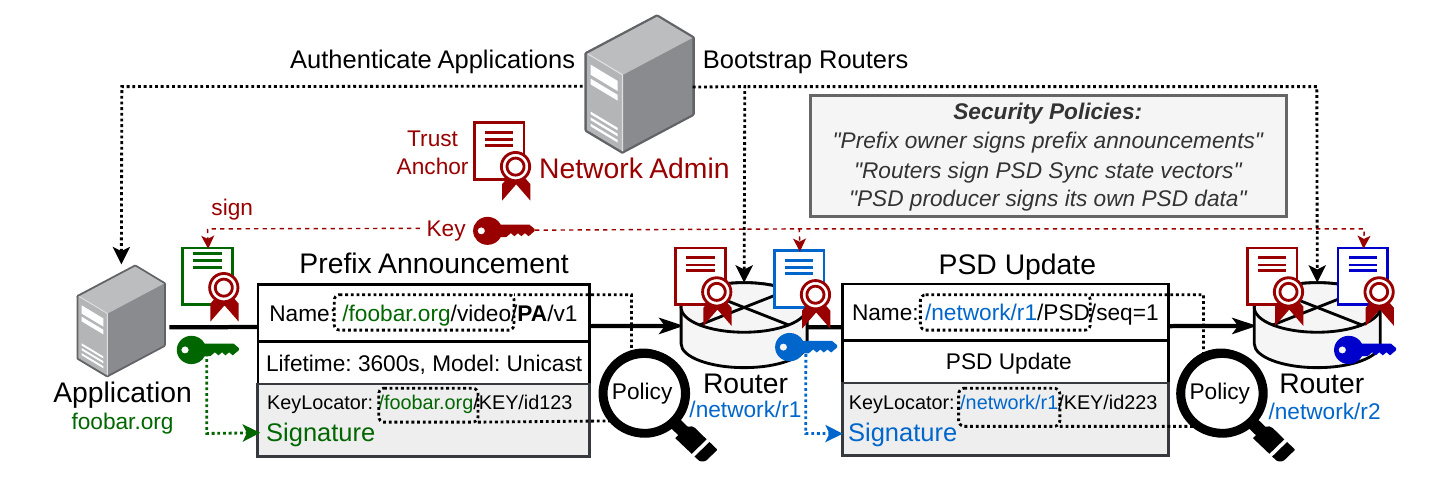}
    \caption{Applications register prefixes with a user-facing router through Prefix Announcements, while the routers secure prefix-to-router mapping state through signed PSD updates. Both verification chains terminate at the network trust anchor, and signing relationships are enforced by security policies.
    PSD Sync state vectors follow a similar naming convention with PSD updates and are therefore omitted for clarity.}
\label{fig:security}
\end{figure*}

\subsection{Maintaining Prefix-to-Router Mappings}
\label{sec:design:psd}
The Prefix State Database (PSD) is a replicated router-to-prefix mapping from which the PET is derived, updated by each user-facing router.

Each subentry describes a prefix reachability object, including its Time-To-Live (TTL), delivery type (unicast or multicast), and local face identifiers (FIDs) through which local applications are attached.
Applications announce their prefixes to their local user-facing router.
After validating the announcement according to the security policy in Section~\ref{sec:design:sec}, the router updates its local PSD and materializes the corresponding PET entry.
Remote routers learn the same prefix reachability state through PSD synchronization and update their own PETs accordingly.

\noindent\textbf{PSD Synchronization:}
To maintain a consistent PET view, SRAN synchronizes PSD instances among user-facing routers.
We view PSD as an ordinary multi-party NDN application that uses the NDN Sync group \name{/<network>/PSD/Sync} to inform the group of the latest named secured data publication.
Each user-facing router publishes local PSD updates as Data objects following the naming convention \name{/<network>/PSD/<router>/<seq>}.
Here, \name{<network>} is the configured network name, while \name{<router>} and \name{<seq>} identify the producing router and its PSD update sequence number.
As a result, each PSD Sync Interest carrying the Sync state vector is conservatively BIER forwarded toward all routers, avoiding the need to separately maintain a user-facing-router membership set.
Routers then forward these BIER packets statelessly using their BIFT.
User-facing routers react to the latest state vector by expressing an Interest based on the naming convention to fetch the missing PSD update, while the router reachability already pre-establishes the forwarding path for this Interest.
The evaluation in Section~\ref{sec:eval} further shows that this synchronization mechanism significantly reduces communication overhead by avoiding network broadcast on new prefix announcement and migration.

\subsection{Routing System Security}
\label{sec:design:sec}

The objective of the PSD security design is to prevent prefix hijacking by external attackers that inject false prefix state into the PSD and by applications that announce prefixes they do not control.
The threat model assumes a valid local trust anchor owned by the network operator and does not cover active attacks against routing protocols or denial-of-service flooding by otherwise authorized entities.

An SRAN network administrator defines a trust domain by self-signing a certificate for the network name as the system trust anchor.
The administrator bootstraps each router with the trust anchor, a router certificate signed by the anchor, and the following security policies: (1) the owners of the DNS name $N$ are authorized to announce prefixes under \name{/N}; for example, owners of \name{foobar.org} can announce \name{/foobar.org/video}; (2) only routers are authorized to produce PSD Sync state vectors; and (3) PSD updates can only be legitimately signed by the same router.
Thus, non-router entities cannot send legitimate PSD Sync Interests, and router \name{/network/r1} cannot produce Data under \name{/network/r2/PSD}.
Fig.~\ref{fig:security} illustrates a scenario in which the administrator of network \name{/network} sets up a network trust anchor by self-signing the network certificate and bootstraps two routers, \name{/network/r1} and \name{/network/r2}.
To announce a prefix to the network, an endpoint application, such as the \name{/foobar.org} server, must be authenticated by the network administrator and obtain a certificate from the trust anchor.

Applications announce their prefixes to user-facing routers using \emph{PrefixAnnouncement} Data named according to convention \name{/<prefix>/PA/<version>}.
The content includes the prefix lifetime (\ie the TTL in PSD) and the delivery model.
As illustrated in Fig.~\ref{fig:security}, the owner of the DNS name \name{foobar.org} has been authenticated and certified by the network administrator and wants to announce a unicast prefix \name{/foobar.org/video} to the network.
It produces announcement Data with the corresponding content and signs it with the anchor-signed \name{foobar.org} key.
Upon receiving the announcement, the router \name{/network/r1} validates the signature against the public key of the trust anchor, then checks that the announcement name is authorized by its Data key locator.
This verification ensures that user-facing routers accept prefix announcements only from authorized producers.
Afterward, the router updates its local PSD and produces signed PSD Data with its certificate.
The strict PSD update signing policy ensures that even if the key \name{/network/r1/KEY/id223} is compromised, the attacker cannot pretend to be another router.

%% file: eval.tex
\section{Implementation and Evaluation}
\label{sec:eval}
In this section, we first describe an NDNd prototype that adds PSD/PET state management, BIER forwarding, and security implementation.
We then measure whether SRAN preserves prefix-update convergence, reduces dissemination overhead, and keeps forwarding state tied to topology rather than application-prefix scale..
Finally, we qualitatively analyze whether the designed routing security mechanisms achieve their design objectives.

\subsection{Implementation}
We implemented SRAN in NDNd, a Go-based NDN forwarder, by extending the forwarder and routing daemon while remaining backward-compatible with applications\rev{\footnote{\url{https://github.com/named-data/ndnd/tree/dv2}}}.
SRAN uses NDNLP as the attachment point: one extension field carries the unicast egress-router identifier, and another carries the BIER bitstring for multicast.
Egress routers are represented by their router names.
Security is enforced using Light VerSec (LVS) trust-schema policies~\cite{yu2015schematizing,yu2023new}: applications announce prefixes via \name{/localhop/route} with signed PrefixAnnouncement Data packets, which are validated against the trust schema before the corresponding prefix state is accepted into the PSD.
The routing daemon implements a PSD module that publishes prefix updates as signed Data objects according to the LVS policies.
Updates are disseminated via BIER-forwarded Sync Interests; user-facing routers apply received updates to their local PSD and populate the PET.
We also implemented management commands that allow operators to inspect and update PSD entries.
BIER forwarding is implemented with a BIFT module in the forwarder, where BIFT entries store the topology-derived next-hop bit masks used by BIER forwarding.
BFR-IDs are assigned via management commands, and the BIFT is automatically rebuilt after FIB updates to remain consistent with the unicast best paths.

\begin{figure}[t]
    \centering
    \includegraphics[width=0.48\textwidth]{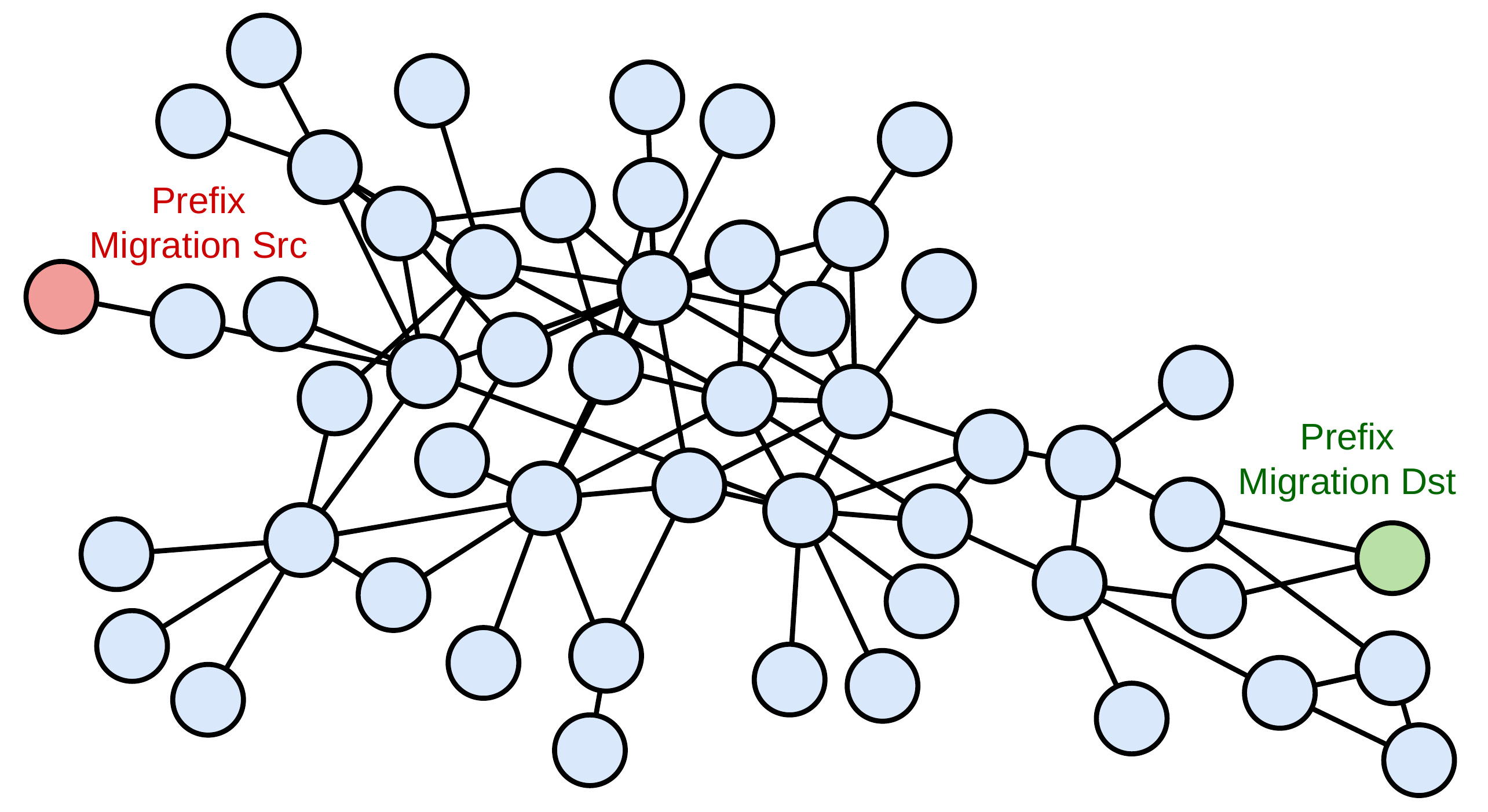}
    \caption{Sprint PoP topology with prefix migration between the farthest routers.}
    \label{fig:sprint}
\end{figure}

\subsection{Performance Evaluation}
\label{sec:eval:eval}
This section evaluates the scalability consequences of moving the intra-domain routing architecture to SRAN. We distinguish three types of state in the evaluation.
The FIB and BIFT constitute \emph{forwarding state}, whose size is determined by network topology at every router.
The PET contains \emph{prefix-to-router mapping state}, which scales with application prefixes and is maintained at user-facing routers. 
When evaluating storage cost, we also report \emph{total router state} as the sum of FIB, BIFT, and PET entries.
The evaluation therefore asks two questions. First, can SRAN maintain prefix-to-router mappings at user-facing routers with timely updates and low communication overhead, and what mapping-state cost does this impose? Second, does SRAN keep forwarding state at all routers bounded by network topology as the number of application prefixes grows?

Throughout the evaluation, we assume a fixed set of user-facing routers (UF routers) for each network topology, and the rest of the routers in the same topology are non-user-facing (NUF routers).
We use two topologies to separate these effects.
Sprint PoP is an all-UF topology where all routers are application-facing.
This is the least favorable case for state reduction, but it still tests whether BIER-based PSD dissemination reduces update overhead and whether hop-by-hop FIB lookup becomes smaller.
Rocketfuel AS~1755 topology~\cite{spring2002rocketfuel} has 61 user-facing routers and 111 non-user-facing routers (NUF routers), and serves as the second evaluation topology.
Both topologies use NDN Distance Vector Routing~\cite{patil2024poster} to establish the router reachability.
The baseline is the existing NDNd forwarder and routing daemon, configured with the same topology and link settings as SRAN.
The key difference is that the baseline installs application prefixes directly into the FIB at every router, whereas SRAN stores application-prefix mappings in the PET at user-facing routers and keeps the hop-by-hop FIB topology-bound.

The measurements use two execution environments, chosen according to topology size.
Topologies smaller than Rocketfuel AS~1755 are emulated with Mini-NDN~\cite{minindn} on the server summarized in Tab.~\ref{tab:eval-platform}.
The AS~1755 prefix-change experiments are simulated with ns-3~\cite{ns3}.
Router state snapshots, comprising the FIB and BIFT \emph{forwarding} tables and the PET \emph{mapping} table, are collected from the corresponding runs.
Finally, to show that the state trend is not specific to AS~1755, we analytically repeat the state calculation on two larger Rocketfuel topologies.

\begin{table}[t]
\centering
\small
\resizebox{\columnwidth}{!}{%
\begin{tabular}{ll}
\toprule
Component & Details \\
\midrule
OS & Ubuntu Linux 5.15.0-171-generic (x86\_64) \\
CPU & AMD EPYC 7702P 64-Core Processor \\
CPU topology & 64 cores / 128 threads, 1 socket, 2 threads per core \\
CPU frequency & 1500--2183 MHz \\
Memory & 251 GB RAM \\
Storage & 776 GB NVMe SSD root volume, 22 TB home volume \\
GPU & ASPEED Graphics Family (AST2500) \\
Kernel/toolchain & Linux 5.15.0-171-generic, GCC 11.4.0 \\
\bottomrule
\end{tabular}%
}
\caption{Evaluation server used for Mini-NDN emulation.}
\label{tab:eval-platform}
\end{table}

\begin{table}[t]
\centering
\small
\resizebox{\columnwidth}{!}{%
\begin{tabular}{lrrrr}
\toprule
Operation & Min & Median & Q3 (P75) & Max \\
\midrule
\textbf{Announce (SRAN)}  & \textbf{98ms} & \textbf{99ms} & \textbf{107ms} & \textbf{110ms} \\
Announce (Baseline) & 95ms & 97ms & 97ms & 118ms \\
\textbf{Migrate (SRAN)}   & \textbf{114ms} & \textbf{125ms} & \textbf{136ms} & \textbf{401ms} \\
Migrate (Baseline) & 109ms & 112ms & 113ms & 365ms \\
\bottomrule
\end{tabular}
}
\caption{Prefix announcement and migration latency on the Sprint PoP topology.}
\label{tab:operation-latency}
\end{table}

Across the evaluation tables, Q3 (P75) denotes the 75th percentile.
For prefix-change experiments, convergence time is the time until the changed prefix is reachable throughout the domain.
The packet-overhead unit ``p'' denotes one packet traversal over one topology link, averaged over all links during the convergence window.

\noindent\textbf{Prefix-Change Convergence and Overhead (Sprint PoP):}
Sprint PoP represents the all-UF deployment case: every router is application-facing.
This experiment therefore asks whether SRAN's PSD dissemination can reduce traffic even before SRAN gets its main benefit.
As shown in Fig.~\ref{fig:sprint}, the emulated topology contains 52 routers and 84 links, and each link is configured with 1~Gbps bandwidth, 10~ms latency, and 0.1\% packet loss.
For announcement, an application prefix is registered at a randomly selected router.
For migration, the prefix is withdrawn from its original router and immediately re-announced at the farthest router, seven hops away as illustrated in Fig.~\ref{fig:sprint}.
We repeat both the announcement and the migration 50 times for SRAN and for the baseline.
\begin{table}[t]
\centering
\small
\resizebox{\columnwidth}{!}{%
\begin{tabular}{lrrrr}
\toprule
Operation & Min & Median & Q3 (P75) & Max \\
\midrule
\textbf{Announce (SRAN)}    & \textbf{2.43p} & \textbf{2.43p} & \textbf{2.43p} & \textbf{4.02p} \\
Announce (Baseline)  & 4.00p & 4.00p & 4.00p & 4.53p \\
\textbf{Migrate (SRAN)}   & \textbf{4.88p} & \textbf{7.29p} & \textbf{7.29p} & \textbf{7.46p} \\
Migrate (Baseline) & 8.00p & 8.02p & 10.81p & 13.67p \\
\bottomrule
\end{tabular}
}
\caption{Normalized per-link packet overhead for prefix announcement and migration on the Sprint PoP topology.}
\label{tab:as1239-normalized-packet-overhead}
\end{table}
\begin{table}[t]
\centering
\small
\resizebox{\columnwidth}{!}{%
\begin{tabular}{lrrrr}
\toprule
Operation & Min & Median & Q3 (P75) & Max \\
\midrule
\textbf{Announce (SRAN)}    & \textbf{2.44p} & \textbf{2.45p} & \textbf{2.45p}  & \textbf{2.47p} \\
Announce (Baseline)  & 4.00p & 8.65p & 13.31p  & 27.27p \\
\textbf{Migrate (SRAN)}   & \textbf{4.88p} & \textbf{4.90p} & \textbf{4.90p}  & \textbf{4.93p} \\
Migrate (Baseline) & 8.00p & 18.86p & 25.07p & 40.62p \\
\bottomrule
\end{tabular}
}
\caption{Normalized per-link packet overhead for prefix announcement and migration on the AS~1755 topology.}
\label{tab:normalized-packet-overhead}
\end{table}
SRAN trades a small amount of convergence latency for lower dissemination overhead.
Tab.~\ref{tab:operation-latency} shows that SRAN keeps convergence latency close to the baseline.
For announcement, the median increases from 97~ms to 99~ms.
For migration, which includes both withdrawal and re-announcement across the farthest router pair, the median increases from 112~ms to 125~ms.
At the 75th percentile, the increases are 10~ms for the announcement and 23~ms for the migration.
The long tail appears in both designs, suggesting that lossy links and Sync retransmission or timer behavior dominate the rare slow cases.
The important point is that SRAN's lower-overhead dissemination does not require a substantially slower convergence process.

Tab.~\ref{tab:as1239-normalized-packet-overhead} shows where SRAN gains even in this unfavorable all-UF topology.
For announcement, SRAN reduces the median packet overhead from 4.00p to 2.43p, a 39.3\% reduction.
For migration, SRAN reduces the median from 8.02p to 7.29p and Q3 from 10.81p to 7.29p, reductions of 9.1\% and 32.6\%, respectively.
The baseline disseminates prefix changes by broadcasting Sync Interests, which creates duplicate packet traversals on many links.
SRAN instead BIER-forwards the PSD Sync Interest; the packet-carried BitString forms an implicit multicast tree and removes many of those duplicates.

\noindent\textbf{Router State Snapshot (Sprint PoP):}\begin{table}[t]
\centering
\small
\resizebox{\columnwidth}{!}{%
\begin{tabular}{lrrrrrr}
\toprule
\multirow{2}{*}{Router} &
\multicolumn{2}{c}{FIB Entries} &
\multicolumn{2}{c}{PET Entries} &
\multicolumn{2}{c}{BIFT Entries} \\
\cmidrule(lr){2-3}\cmidrule(lr){4-5}\cmidrule(lr){6-7}
& Min & Max & Min & Max & Min & Max \\
\midrule
\textbf{SRAN} & \textbf{52} & \textbf{52} & \textbf{1260} & \textbf{1273} & \textbf{52} & \textbf{52} \\
Baseline & 1260 & 1273 & \multicolumn{2}{c}{N/A} & \multicolumn{2}{c}{N/A} \\
\bottomrule
\end{tabular}%
}
\caption{Router state comparison between SRAN and Baseline on the Sprint PoP topology.}
\label{tab:sprint-table-size}
\end{table}
\begin{table}[t]
\centering
\small
\setlength{\tabcolsep}{6pt}
\resizebox{\columnwidth}{!}{%
\begin{tabular}{lrrrrrr}
\toprule
\multirow{2}{*}{Router} &
\multicolumn{2}{c}{FIB Entries} &
\multicolumn{2}{c}{PET Entries} &
\multicolumn{2}{c}{BIFT Entries} \\
\cmidrule(lr){2-3}\cmidrule(lr){4-5}\cmidrule(lr){6-7}
& Min & Max & Min & Max & Min & Max \\
\midrule
\textbf{SRAN NUF Routers} & \textbf{172} & \textbf{172} & \textbf{8} & \textbf{22} & \textbf{172} & \textbf{172} \\
\textbf{SRAN UF Routers} & \textbf{172} & \textbf{172} & \textbf{1380} & \textbf{1388} & \textbf{172} & \textbf{172} \\
Baseline & 1380 & 1388 & \multicolumn{2}{c}{N/A} & \multicolumn{2}{c}{N/A} \\
\bottomrule
\end{tabular}
}
\caption{Router state comparison between SRAN and baseline on the full 172-router Rocketfuel AS~1755 topology with 1000 unicast and 200 multicast application prefixes.}
\label{tab:eval-rf1755-forwarding-state}
\end{table}
Even in an all-UF topology, SRAN should still shrink the table consulted at each hop.
We randomly announce 1000 unicast prefixes and 200 multicast prefixes; each multicast prefix is attached to three egress routers.
In the baseline, all 1200 application prefixes are installed into each router's FIB, together with 60--73 router prefixes.
In SRAN, those application prefixes move to the PET.
The FIB therefore contains only the 52 router-level entries, and the BIFT contains one entry per BFR-ID.

This shift matters because the FIB is consulted at every hop, while the PET is consulted only when performing the initial mapping or local delivery.
Even in the all-UF topology, SRAN reduces the hop-by-hop lookup table from 1260--1273 entries to 52 entries, a 95.9\% reduction.
The PET still contains 1260--1273 entries because every router is user-facing in this topology.
The non-application entries come from distance-vector prefixes and PSD prefixes.
Their small variation reflects the router's degree, because the densely connected routers have more local router prefixes directly from the neighbors.


\noindent\textbf{Prefix-Change Convergence and Overhead (AS~1755):}
We repeat the prefix announcement and migration experiments in AS~1755 using ns-3 simulation.
This topology contains 172 routers. We designate the 111 inferred backbone routers as NUF routers and the remaining 61 as UF routers.
We also configure each link with 1~Gbps bandwidth, 10~ms latency and the same 0.1\% packet loss. 
Announcement and migration both converge in 133~ms in this simulation, so the main difference is the packet overhead.
Tab.~\ref{tab:normalized-packet-overhead} shows that SRAN's BIER dissemination becomes more valuable as the topology grows.
For announcement, SRAN reduces the median overhead from 8.65p to 2.45p and Q3 overhead from 13.31p to 2.45p.
For migration, SRAN reduces the median overhead from 18.86p to 4.90p and Q3 overhead from 25.07p to 4.90p.
The reason is the same as in Sprint PoP, but the effect is greater: the baseline broadcast creates more duplicate traversals in a larger network, while SRAN's BIER approach keeps the dissemination close to a multicast tree.

\noindent\textbf{Router State Snapshot (AS~1755):}
AS~1755 is where SRAN's state reduction should pay off: the user-facing routers maintain prefix-to-router mappings, while the remaining routers do not.
Tab.~\ref{tab:eval-rf1755-forwarding-state} shows the state placement under the same workload of 1000 unicast and 200 multicast prefixes.
The baseline stores 1380--1388 FIB entries at every router: 1200 application prefixes plus 180--188 router prefixes.
SRAN keeps the hop-by-hop FIB fixed at 172 router entries and keeps the BIFT fixed at 172 entries.
The only table whose size depends on application prefixes is the PET, and its size depends on whether the router is user-facing.

This is the core SRAN tradeoff.
Non-user-facing routers keep only 8--22 PET entries, all coming from distance vector routing and PSD Sync.
As a result, the table consulted by non-user-facing routers at each hop drops from 1380--1388 mixed FIB entries to 172 router-level FIB entries, a reduction of up to 87.6\%.
For the unicast lookup path, even if we conservatively add FIB and PET entries together, SRAN NUF routers store only 180--194 entries.
UF routers pay the mapping cost on that path: their PET contains the application-prefix mappings, so their combined FIB and PET state is 1552--1560 entries, exactly 172 entries more than the baseline.

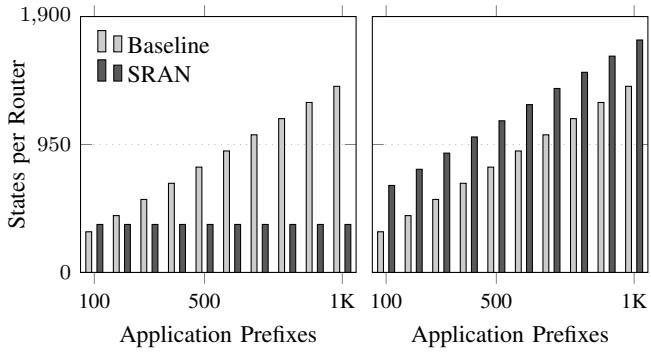
\begin{figure}[t]
\captionsetup{font=small}
\captionsetup[subfigure]{font=small}
\centering
\begin{subfigure}[t]{0.49\columnwidth}
\centering
\begin{tikzpicture}
\begin{axis}[
  ybar,
  scale only axis,
  width=0.84\linewidth,
  height=0.78\linewidth,
  bar width=2.2pt,
  xlabel={Application Prefixes},
  ylabel={States per Router},
  xmin=50,
  xmax=1050,
  ymin=0,
  ymax=1900,
  xtick={100,500,1000},
  xticklabels={100,500,1K},
  ytick={0,950,1900},
  ymajorgrids,
  grid style={dotted},
  tick label style={font=\footnotesize},
  label style={font=\small},
  ylabel style={font=\small,at={(axis description cs:-0.15,0.5)},anchor=south},
  legend style={font=\small,at={(0.03,0.97)},anchor=north west,draw=none,fill=none,row sep=-1pt},
  legend cell align={left},
]
\addplot+[draw=black,fill=black!20] coordinates {
  (100,301.918) (200,421.918) (300,541.918) (400,661.918) (500,781.918)
  (600,901.918) (700,1021.918) (800,1141.918) (900,1261.918) (1000,1381.918)
};
\addplot+[draw=black,fill=black!65] coordinates {
  (100,356.2613) (200,356.2613) (300,356.2613) (400,356.2613) (500,356.2613)
  (600,356.2613) (700,356.2613) (800,356.2613) (900,356.2613) (1000,356.2613)
};
\legend{Baseline,SRAN}
\end{axis}
\end{tikzpicture}
\end{subfigure}%
\hfill
\begin{subfigure}[t]{0.49\columnwidth}
\centering
\begin{tikzpicture}
\begin{axis}[
  ybar,
  scale only axis,
  width=0.84\linewidth,
  height=0.78\linewidth,
  bar width=2.2pt,
  xlabel={Application Prefixes},
  xmin=50,
  xmax=1050,
  ymin=0,
  ymax=1900,
  xtick={100,500,1000},
  xticklabels={100,500,1K},
  ytick={0,950,1900},
  yticklabels=\empty,
  ymajorgrids,
  grid style={dotted},
  tick label style={font=\footnotesize},
  label style={font=\small},
  title style={font=\small},
]
\addplot+[draw=black,fill=black!20] coordinates {
  (100,301.918) (200,421.918) (300,541.918) (400,661.918) (500,781.918)
  (600,901.918) (700,1021.918) (800,1141.918) (900,1261.918) (1000,1381.918)
};
\addplot+[draw=black,fill=black!65] coordinates {
  (100,645.9180) (200,765.9180) (300,885.9180) (400,1005.9180) (500,1125.9180)
  (600,1245.9180) (700,1365.9180) (800,1485.9180) (900,1605.9180) (1000,1725.9180)
};
\end{axis}
\end{tikzpicture}
\end{subfigure}%
\caption{Router state as application prefixes increase in AS~1755. Bars compare the baseline against total SRAN router state, counted as the sum of FIB, PET, and BIFT entries. The left plot shows NUF routers; the right plot shows UF routers.}
\label{fig:eval-rf1755-lookup-sweep}
\end{figure}

\noindent\textbf{The Effect of Application Dynamics:}
The previous snapshot shows one workload size.
We next hold the AS~1755 topology fixed and increase the number of application prefixes.
If SRAN works as intended, the baseline NUF FIB should grow with every new application prefix, while the SRAN NUF FIB should remain flat.

We sweep the number of unicast prefixes $U$ from 100 to 1000 in steps of 100 and set the multicast-prefix count to $U/5$.
Thus, the final sweep point matches the workload in Tab.~\ref{tab:eval-rf1755-forwarding-state}.
For the baseline, each router stores $U+U/5$ application-prefix FIB entries plus router prefix state.
We use 181.92 entries per router for the router component, corresponding to the average of the 180--188 range in Tab.~\ref{tab:eval-rf1755-forwarding-state}.
For SRAN, we conservatively count the total local router state as the sum of FIB, PET, and BIFT entries, although BIFT and FIB are never used at the same time in Interest forwarding.

Fig.~\ref{fig:eval-rf1755-lookup-sweep} shows the expected split.
The baseline grows linearly from 301.92 entries per router at $U=100$ to 1381.92 entries at $U=1000$.
In contrast, each SRAN NUF router remains fixed at 356.26 entries: 172 FIB entries, 12.26 PET entries on average, and 172 BIFT entries.
This fixed cost is 18.0\% higher than the baseline at $U=100$, but it does not grow with the application workload.
At $U=1000$, SRAN reduces total router state by 74.2\% in NUF routers.

SRAN intentionally keeps application-prefix mappings in the PET, so UF router state grows with the same slope as the baseline.
The average SRAN UF router state increases from 645.92 entries at $U=100$ to 1725.92 entries at $U=1000$.
The difference from baseline is a fixed 344-entry offset from the router-level FIB and BIFT tables maintained alongside the user-facing PET.
Thus, SRAN does not remove prefix state from the network; it moves that state to the UF routers and keeps the extra cost topology-bounded.
The BIFT remains fixed at 172 entries throughout the sweep, showing that multicast forwarding state is bounded by router reachability rather than by the number of multicast prefixes.

\noindent\textbf{The Effect of Physical Topologies:}
Finally, we ask whether the same trend holds as the physical topology changes.
The previous sweep varies application scale on one topology; this sweep asks how topology size shifts the point where SRAN's fixed topology cost is outweighed by prefix growth.
Because the forwarding tables are determined by topology and advertised prefixes, we conduct an analytical state-scaling study on three Rocketfuel topologies: AS~1755, AS~3356, and AS~2914.
AS~3356 and AS~2914 contain 624 and 960 routers, respectively.
We vary the application workload from 1000 to 10{,}000 prefixes announced at the UF routers and measure two quantities: the percentage of state that SRAN removes from NUF routers, and the additional state that SRAN introduces at the UF routers.
If NUF router state is topology-bound, then as the prefix workload grows, the state reduction should approach 100\%.
Because the extra cost is fixed by the topology, its relative overhead should approach 0\%.

\begin{figure}[t]
\captionsetup{font=small}
\captionsetup[subfigure]{font=small}
\centering
\begin{subfigure}[t]{0.48\columnwidth}
\centering
\begin{tikzpicture}
\begin{axis}[
  scale only axis,
  width=0.84\linewidth,
  height=0.78\linewidth,
  xlabel={Application Prefixes},
  ylabel={Percentage},
  xmin=0.8,
  xmax=10.2,
  ymin=0,
  ymax=100,
  xtick={1,5,10},
  xticklabels={1K,5K,10K},
  ytick={0,20,40,60,80,100},
  ymajorgrids,
  grid style={dotted},
  tick label style={font=\footnotesize},
  label style={font=\small},
  ylabel style={font=\small,at={(axis description cs:-0.1,0.5)},anchor=south},
  legend style={font=\small,at={(0.97,0.03)},anchor=south east,draw=none,fill=none,row sep=-1pt},
  legend cell align={left},
]
\addplot+[draw=black,mark=*,thick] coordinates {
  (1,69.86) (2,83.67) (3,88.80) (4,91.48) (5,93.12)
  (6,94.24) (7,95.04) (8,95.65) (9,96.12) (10,96.50)
};
\addplot+[draw=black,mark=square*,thick,dashed] coordinates {
  (1,21.97) (2,51.58) (3,64.90) (4,72.47) (5,77.36)
  (6,80.77) (7,83.29) (8,85.22) (9,86.76) (10,88.00)
};
\addplot+[draw=black,mark=triangle*,thick,dotted] coordinates {
  (1,1.74) (2,34.84) (3,51.25) (4,61.06) (5,67.59)
  (6,72.24) (7,75.72) (8,78.43) (9,80.59) (10,82.36)
};
\legend{AS~1755,AS~3356,AS~2914}
\end{axis}
\end{tikzpicture}
\end{subfigure}%
\hfill
\begin{subfigure}[t]{0.48\columnwidth}
\centering
\begin{tikzpicture}
\begin{axis}[
  scale only axis,
  width=0.84\linewidth,
  height=0.78\linewidth,
  xlabel={Application Prefixes},
  xmin=0.8,
  xmax=10.2,
  ymin=0,
  ymax=100,
  xtick={1,5,10},
  xticklabels={1K,5K,10K},
  ytick={0,20,40,60,80,100},
  yticklabels=\empty,
  ymajorgrids,
  grid style={dotted},
  tick label style={font=\footnotesize},
  label style={font=\small},
  legend style={font=\small,at={(0.97,0.53)},anchor=south east,draw=none,fill=none,row sep=-1pt},
  legend cell align={left},
]
\addplot+[draw=black,mark=*,thick] coordinates {
  (1,29.11) (2,15.77) (3,10.81) (4,8.23) (5,6.64)
  (6,5.56) (7,4.79) (8,4.20) (9,3.75) (10,3.38)
};
\addplot+[draw=black,mark=square*,thick,dashed] coordinates {
  (1,76.31) (2,47.35) (3,34.33) (4,26.92) (5,22.15)
  (6,18.81) (7,16.34) (8,14.45) (9,12.95) (10,11.73)
};
\addplot+[draw=black,mark=triangle*,thick,dotted] coordinates {
  (1,97.50) (2,64.66) (3,48.37) (4,38.64) (5,32.17)
  (6,27.55) (7,24.09) (8,21.41) (9,19.26) (10,17.50)
};
\legend{AS~1755,AS~3356,AS~2914}
\end{axis}
\end{tikzpicture}
\end{subfigure}
\caption{Analytical state scaling across Rocketfuel topologies. State reduction (left) measures the percentage of baseline entries removed by SRAN at the NUF routers. State overhead (right) measures the additional SRAN UF state relative to the baseline.}
\label{fig:eval-theoretical-state-curves}
\end{figure}
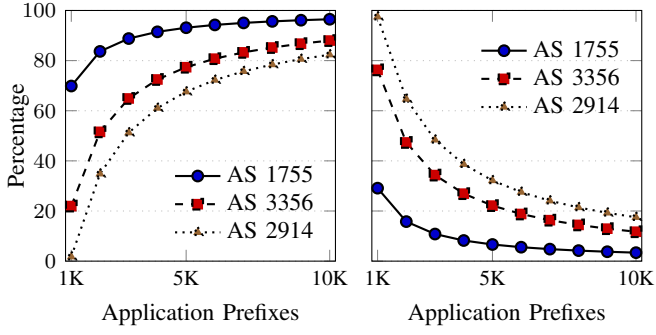

Fig.~\ref{fig:eval-theoretical-state-curves} reports the analytical results, using the same criterion of counting the combined FIB, PET, and BIFT entries as router state.
Across all three topologies, the state reduction increases as the application workload grows.
The baseline FIB in NUF routers absorbs every new application prefix, whereas SRAN FIBs, PETs, and BIFTs in those routers remain topology-bound.
As the number of prefixes increases from 1000 to 10{,}000, the total state reduction in NUF routers increases from 69.9\% to 96.5\% on AS~1755, from 22.0\% to 88.0\% on AS~3356, and from 1.7\% to 82.4\% on AS~2914.
All three curves trend toward 100\%, confirming that the SRAN NUF router state is bounded by the physical topology rather than by the application namespace.
Larger topologies start with a lower percentage reduction because they impose larger topology-derived SRAN state; they therefore require a larger prefix workload before the relative reduction approaches 100\%.
At the UF routers, SRAN adds a fixed topology-derived overhead of 344, 1248, and 1920 entries per router on AS~1755, AS~3356, and AS~2914, respectively.
Because this overhead is fixed by the router set, its relative cost decreases as the application workload grows and trends toward 0\%.
\subsection{Security Analysis}
\label{sec:eval:security}

We qualitatively analyze the routing security design from Section~\ref{sec:design:sec}.
The security question is a state-admission question: can an attacker make a user-facing router install false prefix state?
There are two admission points for such state: application PrefixAnnouncements and router-produced PSD updates.
We use the same trust model as the design: routers are certified under the network trust anchor, and network administrators issue certificates to application prefix owners.

\noindent\textbf{Unauthorized Prefix Announcement:}
An attacker may try to register a victim prefix, such as \name{/foobar.org/video}, without owning \name{/foobar.org}.
Before installing PSD or PET state, the user-facing router validates the \emph{PrefixAnnouncement} Data with the bootstrapped LVS trust schema.
The schema checks both the certificate chain to the network anchor and the authorization relationship between the announced prefix and the signing key.
An unsigned announcement, an announcement signed by a key from another namespace, or an announcement that over-claims beyond the signer's delegated namespace is rejected before it changes router state.

\noindent\textbf{Forged PSD Update:}
An attacker may inject fabricated or replayed PSD Data, for example under \name{/network/r2/PSD}.
Receiving user-facing routers accept the update only if it is signed by the router authorized for that PSD name and carries a newer sequence number than the current state.
Thus, non-router keys, cross-router signatures, and stale sequence numbers are rejected before they update the PET; prefix lifetimes further limit the effect of replayed announcements.

\noindent\textbf{Compromised Router Key:}
If \name{/network/r1}'s key is compromised, the attacker can publish state only under \name{/network/r1/PSD}.
It cannot impersonate \name{/network/r2} or withdraw prefixes on behalf of other routers.
Other user-facing routers, therefore, reject cross-router misinformation even when it is disseminated through normal Sync and BIER forwarding.


%% file: related.tex
\section{Related Work}
\label{sec:related}

\subsection{Routing Scalability}

\noindent\textbf{Map-and-Encap:}
Map-and-Encap realizes the routing-scalability principle of mapping external identifiers to topology-derived routing identifiers. 
LISP maps the logical endpoint prefixes to globally routable locators~\cite{rfc9300}, and MPLS maps forwarding equivalence classes to labels~\cite{rfc3031}.
SRAN adopts the same principle but uses \emph{Map-and-Attach}: it maps name prefixes to egress routers and attaches the mapping without encapsulating the Interest packet as an opaque payload.
As a result, NDN's data-centric delivery semantics are preserved.

\noindent\textbf{Hyperbolic Routing:}
Hyperbolic routing maps topological entities such as routers into a hyperbolic coordinate space and forwards toward a destination coordinate~\cite{kleinberg2007geographic,papadopoulos2010greedy,
papadopoulos2015network,cvetkovski2009hyperbolic,
shavitt2008hyperbolic,lehman2016hyperbolic}. 
It therefore addresses scalable route computation/forwarding over topology, whereas SRAN addresses mapping the application namespace onto those topological destinations. The two are orthogonal and complementary.  For example, SRAN could supply the prefix-to-router mapping while hyperbolic routing forwards toward the mapped router.

\noindent\textbf{FIB Aggregation:}
FIB aggregation~\cite{draves1999constructing,zhao2010aggregatability,
ballani2009viaggre,uzmi2011smalta,retvari2013compressing} reduces forwarding state by combining structured destinations that have compatible next hops.
It can be applied to IP prefixes and semantic names.
However, aggregation retains these destinations inside the routing and forwarding system, and its effectiveness depends on how well they can be summarized.
SRAN instead removes application prefixes from the FIB altogether, mapping them to topology-bound router identifiers before hop-by-hop forwarding.
The prefix-to-router mappings themselves are maintained at user-facing routers, separately from RIB and FIB.

\subsection{BIER}
BIER itself is not a contribution of SRAN.
Rather, SRAN contributes a distributed mapping system that enables BIER-based forwarding by mapping each application name prefix to a set of egress routers.
More importantly, this mechanism supports both scalable unicast and multicast: a singleton set specifies a unicast destination, whereas a multi-router set specifies multicast destinations.

\subsection{NDN Scalability}
\noindent\textbf{SNAMP:}
SNAMP~\cite{afanasyev2015snamp} is the closest prior NDN work on secure namespace mapping for scalable forwarding. It resolves a data name through NDNS~\cite{afanasyev2017ndns} and carries the returned locator in a newly introduced Interest packet field.
SRAN generalizes this approach by distributing the mapping system across routers, mapping each application name prefix directly to one or more egress-router names, and attaching the mapping result without modifying the original Interest, which remains intact end to end.
This common mapping abstraction supports both unicast and multicast while preserving NDN's native data-centric delivery semantics.

\noindent\textbf{Scalable Name Prefix Lookups:}
Scalable name prefix lookup techniques reduce the memory and processing costs associated with a given prefix set~\cite{songsigcomm25, yuan2017,ghasemi2019triegranularity} and are thus complementary to SRAN by reducing the storage and processing burden at the user-facing routers.

%% file: discuss.tex
\section{Discussion}
\label{sec:discuss}

\subsection{Scalable Routing Architecture}
The architectural principle underlying Map-and-Encap is to map external identifiers onto network topology so that routing and forwarding scale with topological units rather than the external identifier space. SRAN applies this principle to intra-domain NDN through Map-and-Attach, replacing encapsulation with attached forwarding metadata to preserve NDN's data-centric delivery semantics. The principle is independent of the particular routing protocol, mapping-maintenance mechanism, multicast mechanism, or attachment mechanism used by SRAN.

\subsection{Inter-Domain Applicability}
The same routing scalability principle can be applied to inter-domain routing, with the topological routing unit changing from routers to
Autonomous Systems (ASes). Within one domain, SRAN maps an application prefix to one or more egress routers, allowing forwarding to operate on router reachability. 
At the inter-domain level, an application prefix can similarly be mapped to one or more destination ASes, allowing inter-domain forwarding to operate on AS reachability.
After an Interest reaches a destination AS, the intra-domain mapping
described in this paper maps the prefix to the appropriate egress
router. Thus, the mapping principle can be applied hierarchically
without changing the application name carried by the Interest.

Unlike the intra-domain case, however, prefix reachability propagation across ASes is governed by routing policies, making the PSD-style global synchronization of prefix mapping inapplicable.  
In practice, BGP already establishes the prefix reachability propagation model between neighboring ASes based on each AS's local import and export policies, and associates an advertised prefix with its origin AS. 
Therefore, an NDN inter-domain routing protocol can conceptually follow the same policy-driven model to disseminate prefix-to-AS mappings, enabling a destination AS identifier to be attached to each Interest. Upon reaching the destination AS, SRAN can use the intra-domain prefix-to-router mapping to steer the Interest to its destination.

An Internet-wide NDN realization will change the routing-security problem. In an NDN Internet, an inter-domain routing protocol can be implemented as an NDN application, as NLSR and NDN Distance Vector are within a routing domain, and exchange routing updates as named, cryptographically secured Data. 
The remaining challenge is establishing scalable trust across ASes: an AS must be able to authenticate routing Data produced by other ASes
and verify that an AS is authorized to originate a given application
prefix. Today's Internet addresses the analogous prefix-to-origin-AS
authorization problem through RPKI, where resource holders
cryptographically authorize ASes to originate their IP prefixes.
We are currently investigating how existing inter-domain routing and security mechanisms can be adapted to NDN prefix-to-AS mapping.

%% file: conclude.tex
\section{Conclusion}
\label{sec:conclude}

This paper presented SRAN, an intra-domain NDN routing architecture
that maps application name prefixes to topological destinations while
preserving NDN's native data-centric communication model. SRAN realizes
this separation through Map-and-Attach: prefix-to-egress-router mappings
are carried in NDNLP while the original Interests remain intact,
allowing PIT aggregation, in-network caching, and Data return to operate
unchanged. The same mapping supports both Interest unicast forwarding and
BIER-based Interest multicast. Our evaluation of the SRAN prototype shows that routing and forwarding state at routers scales with network topology rather than the application namespace, while prefix-to-router mappings are maintained in real time with low dissemination overhead.

More broadly, SRAN demonstrates that the routing-scalability principle
underlying Map-and-Encap can be applied to NDN without using
encapsulation. By separating application-prefix reachability from
topological reachability, Map-and-Attach allows routing and forwarding
to scale with network topology without impacting NDN's native
data-centric delivery semantics. This architectural separation is
independent of the particular routing, mapping synchronization,
multicast forwarding, and attachment mechanisms used in our prototype,
allowing these mechanisms to evolve independently.